\documentclass[pre,twocolumn,showpacs,floatfix]{revtex4}
\usepackage{graphicx,amsmath,amssymb,url}

\begin{document}
\title{Modeling scientific-citation patterns and other triangle-rich acyclic networks}

\author{Zhi-Xi Wu}
\email{zhi-xi.wu@physics.umu.se}
\author{Petter Holme}
\email{petter.holme@physics.umu.se} \affiliation{Department of
Physics, Ume{\aa}\, University, 901 87 Ume{\aa}, Sweden}

\begin{abstract}
We propose a model of the evolution of the networks of scientific
citations. The model takes an out-degree distribution
(distribution of number of citations) and two parameters as input.
The parameters capture the two main ingredients of the model, the
aging of the relevance of papers and the formation of triangles
when new papers cite old. We compare our model with three network
structural quantities of an empirical citation network. We find
that an unique point in parameter space optimizing the match
between the real and model data for all quantities. The optimal
parameter values suggest that the impact of scientific papers, at
least in the empirical data set we model is proportional to the
inverse of the number of papers since they were published.
\end{abstract}
\maketitle

\section{Introduction}\label{intro}

The boom of networks studies of the last
decade~\cite{Newman2003rev,Boccaletti2006pr} has potentially an
impact of the structure of science itself. Network measures can
help creating better bibliometric quantities to evaluate
scientific impact~\cite{Redner2005phystoday} and the sociological
aspect of scientific collaboration and exchange of ideas. Indeed,
the study of scientific citations has become a subfield of complex
network
studies~\cite{price,Redner1998epjb,Klemm2002pre,Zhu2003pre,
Lehmann2003pre,Sen2005pa,Hajra2005pa,Hajra2006pa,leicht,Wang2008pa}.

\begin{figure}
\includegraphics[width=.7\linewidth]{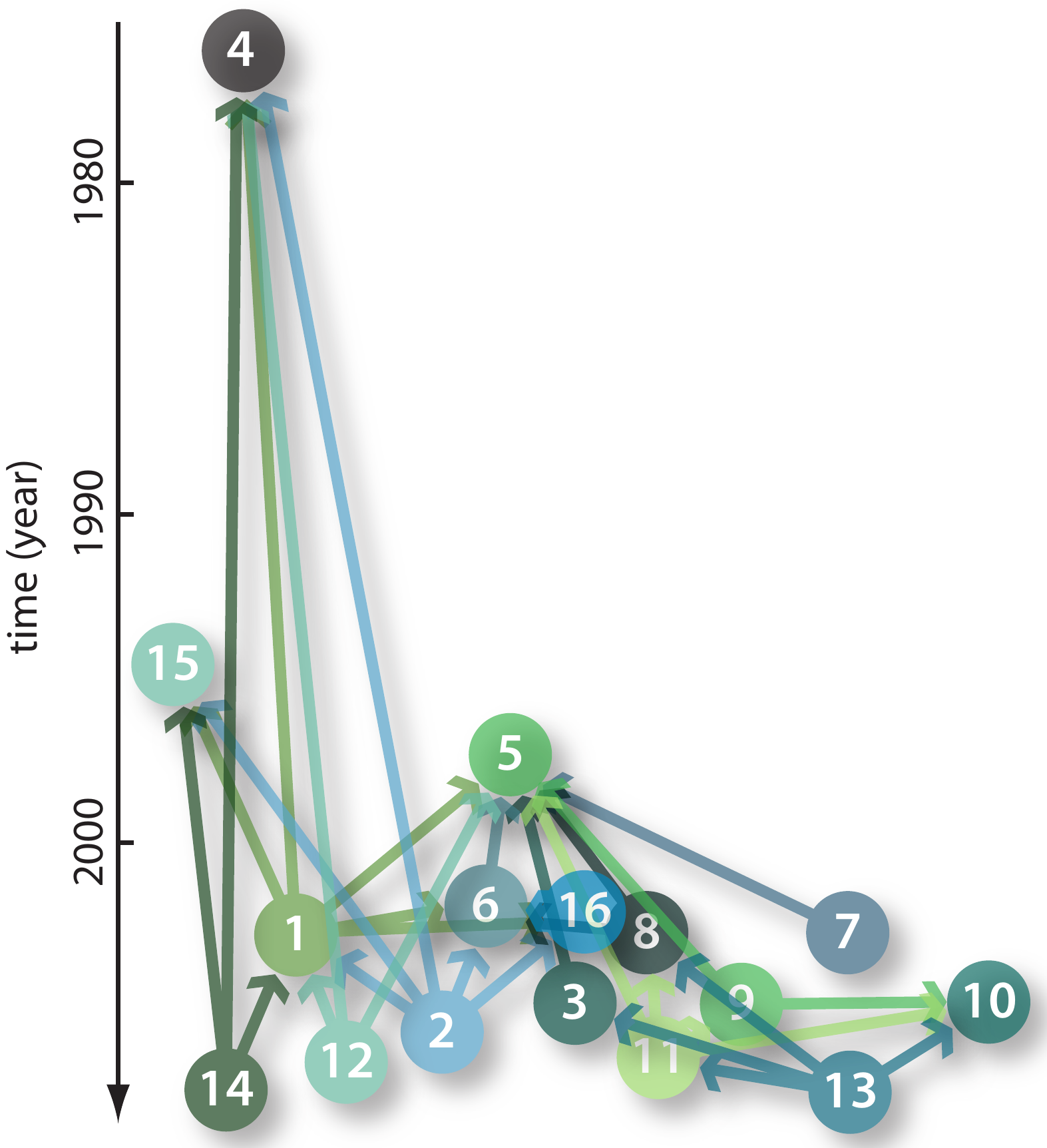}
\caption{(Color online) An example of a citation network --- the
citation network of articles cited by this articles (with the
indices being the indices of the reference list).\label{fig:ex}}
\end{figure}

One typical feature of academic citation networks is that the
number of citations to a paper decreases with its age. Inspired by
this point, many works have been focused on how a paper's age
influences its ability to attract new
citations~\cite{Sen2005pa,Hajra2005pa,Hajra2006pa,Wang2008pa} (or,
equally, new \textit{attachments} in the network). Specifically,
it is believed that the attachment rate (the rate of new citations
to an old paper) is dependent on both the current number of
citations (its \textit{in-degree} in the network) and its age.
(Here we consider citations going back in time meaning that
out-degree is the number of references and in-degree is the number
of citations.) Another important constraint of citation networks
is that they are time ordered --- of any pair of papers, one is
the oldest. (It might, in practice, be more relevant to consider
papers published almost simultaneously unordered, but in this work
we assume this is a negligible effect). An important consequence
of the time ordering is that citation networks are acyclic, i.e.\
there are no closed (directed) paths.  In Fig.~\ref{fig:ex} we
show a small citation network as an example. This network shows is
the references of this paper and how they cite each other. In a
recent paper~\cite{Karrer2009prl}, Karrer and Newman (KN) proposed
a random graph model for directed acyclic graphs. In the KN model,
the vertices are ordered by time and their in- and out-degrees are
pre-assigned (similar to the undirected ``configuration
model''~\cite{Molloy1995rsa}). The vertices are added to the
network iteratively (from $1$ to $N$, with $N$ being the network
size), and for each new vertex $v$, arcs (directed edges) are
added from old vertices whose in-degree is lower than their
prescribed value until $v$'s out-degree is as large as its
prescribed value. Karrer and Newman validate their model with
empirical measurements and get good agreements for some
quantities~\cite{Karrer2009prl}, but their model does, as we will
show, not generate as many triangles as real citation networks
have. (Note that there are two topologically different directed
triangles, but only one of them is acyclic, which makes the word
``triangle'' unique in this study.) In this work, we present a
model of academic citation networks that remedies the lack of
triangles in the KN model by building on mechanisms arguably at
work in the scientific process. In this paper, we first discuss
the structure of empirical citation networks, then present the
model and last test it against three network-structural quantities
of real citation networks.

\begin{figure}
\includegraphics[width=.625\linewidth]{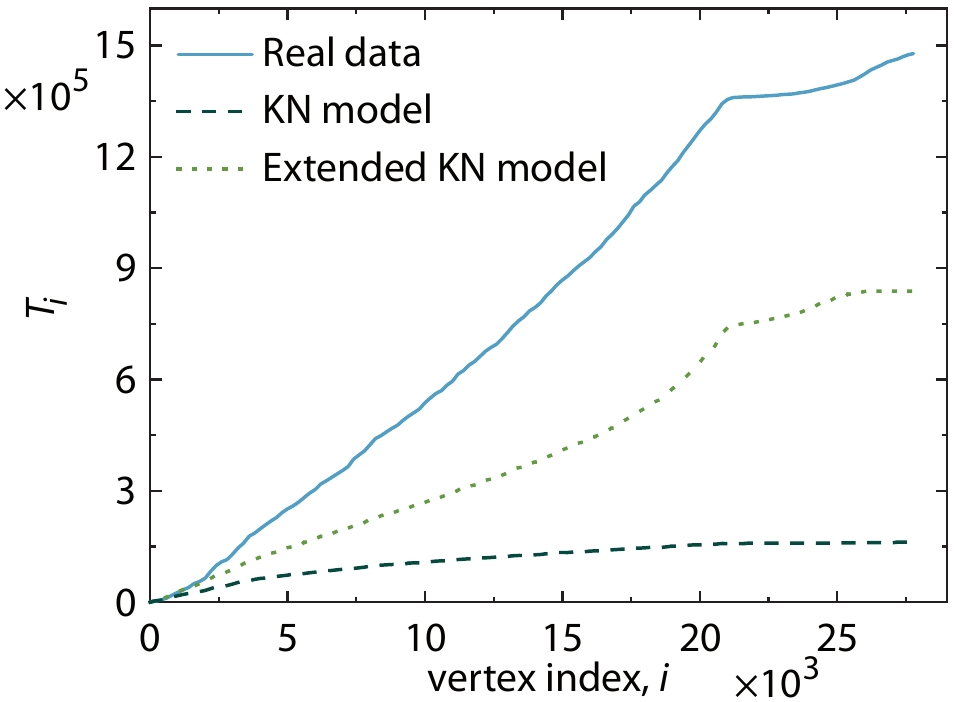}
\caption{(Color online) The number $T_i$ of triangles as a
function of the order the vertex is added $i$. The solid, dotted,
and dashed lines correspond to the empirical citation network of
high-energy physics papers, the KN model and the extended KN
model, respectively. The values of $T_i$ for the two theoretical
models are averages over two hundred independent samples.
\label{fig:kn}}
\end{figure}

\section{Empirical measurements and the predictions of the KN model}

Before presenting our model we state the most important motivation
for this study. In Fig.~\ref{fig:kn} we show the number $T_i$ of
triangles in an empirical citation network consisting of
$N=27{,}770$ papers (or rather preprints) on theoretical
high-energy physics. There are in total $352{,}285$ citations (or
\textit{arcs}, directed edges) among them. The data set comes from
preprints posted on \url{arxiv.org} between 1992 and 2003. By
measurement, we define a triangle as the pattern ``paper $A$
citing $B$ and $C$, and $B$ citing $C$'', and calculate the number
of such patterns present in the network when going through the
papers from $1$ to $N$ (the order of their appearance on the
website). To reduce the computational complexity, we sample each
$200$'th $i$-value. For comparison, we also plot the predicted
number of triangles of the KN model, and a simple extension of the
KN model introducing more triangles: When a new vertex enters the
network, rather than randomly matching all its out-degrees with
those in-degrees among the existing vertices, after first matching
one out-degree randomly with an in-degree belonging to an older
vertex $w$ (like the KN model), we let as many of the remaining
arcs as possible to come from neighbors of $w$ (and after that,
also the neighbors of its new neighbor). Note that, by the
definition of the KN model both the network size $N$ and the
degree sequences (both in- and out-degrees) are identical with the
empirical data. Both the KN model and the extension underestimate
the number of directed triangles in the real network.

\section{Motivation and definition of the model}\label{model}

In this section we will discuss and motivate our model. We start
by ordering the vertices temporally as in the real data, and their
out-degrees (the number of citations) are kept as the same as the
original. (Alternatively the degrees can be drawn from some
appropriate distribution.) We do not restrict the number of
in-degrees --- that will be an emergent property of the model that
we will use for validation. We add the vertices one by one and
fill up the out-degrees of the new vertex before adding a new.

A common assumption is that the relevance of a paper decays with
its age~\cite{Redner2005phystoday,
Redner1998epjb,Klemm2002pre,Zhu2003pre,Sen2005pa,
Hajra2005pa,Hajra2006pa,Karrer2009prl}. In other words, science
will move away from any paper. For this reason, we let the first
arc from a new vertex $i$ go to an old vertex with a probability
$\prod_{i\to j}$ proportional to its age $t_j=i-j$ to a power
$\alpha$ (where a negative $\alpha$ reflect an attachment
probability decaying with age). For to fill up the remaining
out-degrees of $i$, we attach arcs with probability $\beta$ to
random (in- or out-) neighbors of $j$, and otherwise (i.e.\ with
probability $1-\beta$) attach arcs to older vertices with
probability~$\prod_{i\to j}$ as above. If there is no available
neighbor to attach to (we assume one vertex cannot link to another
vertex twice, or to itself), we make an attachment of the first
type. Note that the number of candidates whom $i$ can connect to
increases with more out-degrees in the system, i.e.\ with time.
This \textit{triangle-formation step} (proposed in Ref.~\cite{HK}
as a model of scale-free networks with a tunable clustering
coefficient) is a mechanism that, we argue fits well to citation
networks. To put a scientific paper in the right context one cite
papers of the same theme, since these papers are similar to each
other they are likely to each other. This in itself means that we
can expect many triangles
--- if paper A cites B and C and B also cites C with a relatively
large probability, which is effectively the same as the triangle
formation sketched above. As a more explicit mechanism one can
imagine that when working on paper A the researchers may find
paper C from the reference list of paper B. In sum, our model has
two input parameters $\alpha$ and $\beta$ (in addition to the
degrees), governing the two key ingredients
--- aging and triangle formation.

\section{Measured quantities}
Following Ref.~\cite{Karrer2009prl}, for each vertex $i$, we
define a parameter
\begin{equation}
\lambda_i=\sum_{j=1}^{i-1}k_j^{\mathrm{in}}-\sum_{j=1}^ik_j^{\mathrm{out}}.
\end{equation}
$\lambda_i$ is thus the sum of in-degrees of the vertices that
have been added in the network before $i$ (i.e., from the vertex
$1$ to the vertex $i-1$) minus the sum of in-degrees. As pointed
out in Ref.~\cite{Karrer2009prl}, this parameter should satisfy
the conditions $\lambda_i\ge0$ for $i=2,\cdots, n-1$ and
$\lambda_1=\lambda_n=0$. The interpretation of $\lambda_i$ is that
it is the number of arcs that connecting vertices later than $i$
to vertices earlier than $i$ \cite{Karrer2009prl}. We will also
measure $P(k_{\mathrm{in}})$, the probability of randomly
selecting a vertex whose in-degree is $k_{\mathrm{in}}$, and
$T_i$. After the networks are constructed, we measure these three
quantities and compare them with the corresponding empirical
values. The results presented below for models are averages over
$200$ independent network realizations.

\begin{figure}
\includegraphics[width=\linewidth]{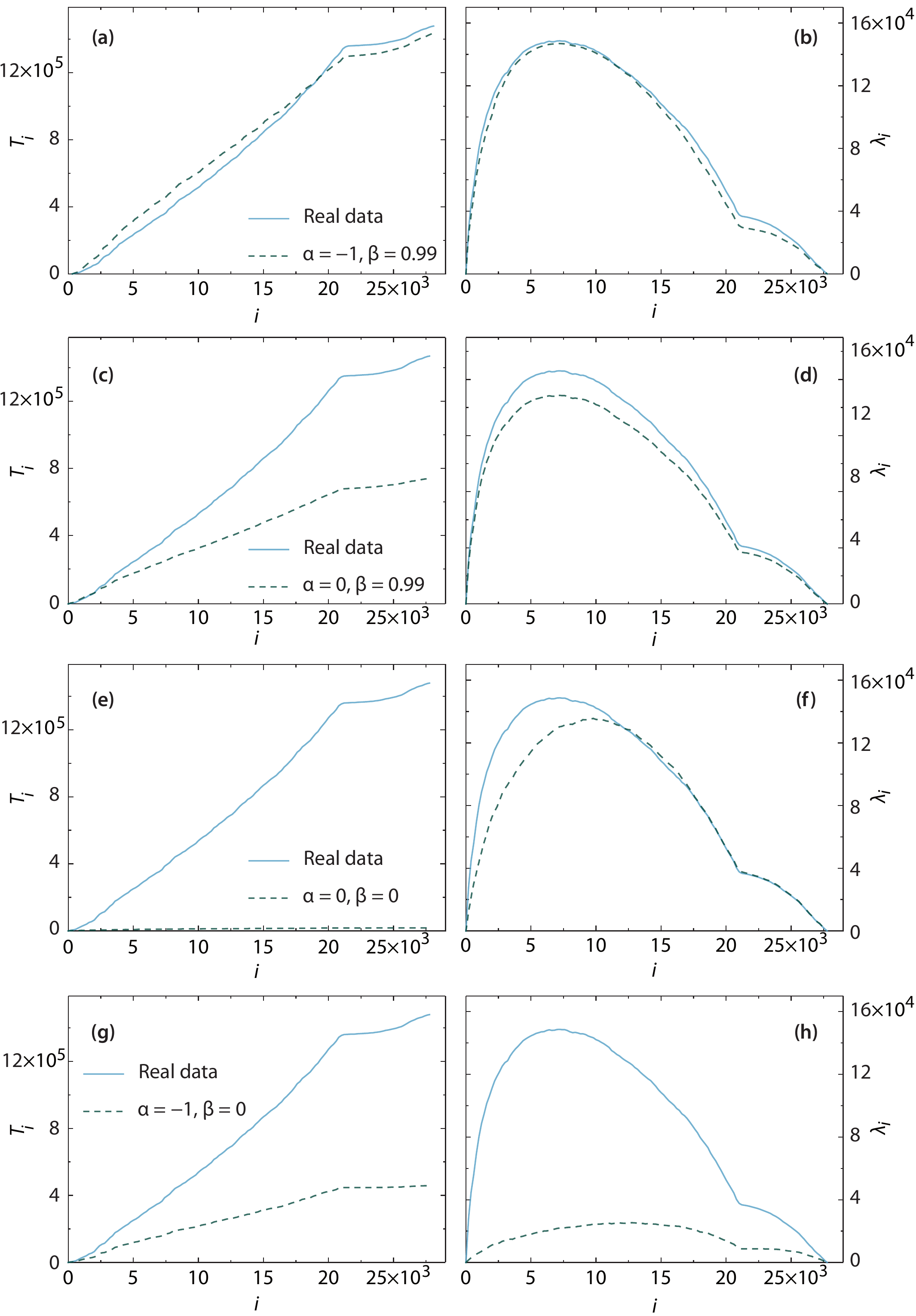}
\caption{(Color online) Network statistics for our model. The
solid lines correspond to the citation network of high-energy
physics papers and the dashed lines represent our model data. In
(a) and (b) we use the model parameters $\alpha=-1$ and
$\beta=0.99$. (a) shows the average number $T_i$ of triangles as a
function of the index of the added vertex $i$.  (b) displays the
number of arcs passing $i$, $\lambda_i$ as a function of $i$. (c)
and (d) corresponds to (a) and (b) but for $\alpha=0$ and
$\beta=0.99$. (e) and (f) show the same for $\alpha=\beta=0$. (g)
and (h) also corresponds to (a) and (b) but for parameters
$\alpha=-1$ and $\beta=0$.\label{fig:our}}
\end{figure}

\section{Results}\label{results}

Now we turn to the numerical results for our model. We first
investigate the model dependence on the parameters $\alpha$ and
$\beta$ and compare the values of $T_i$ and $\lambda_i$ to the
real data. By construction, large $\beta$-values give large
numbers of triangles. As seen in Fig.~\ref{fig:our}(a) there are
(unlike the results in Fig.~\ref{fig:kn}) parameters giving a
number of triangles that matches the empirical curves. A negative
$\alpha$-value is important, not only to get $T_i$-values matching
the empirical data, but also to obtain matching $\lambda_i$-values
(Fig.~\ref{fig:our}(b)).  We have scanned the region of
$\alpha\in[-2, 0]$ and $\beta\in[0, 1]$, and found that the
combination $\alpha=-1$ and $\beta=0.99$ gives the best fit to the
empirical data~\cite{Notes}. To give an overview of the model's
behavior we plot three other combinations of $\alpha$- and
$\beta$-values in Fig.~\ref{fig:our}. In Fig.~\ref{fig:our} (c)
and (d) we show the results for $\alpha=0$ and $\beta=0.99$. When
$\alpha=0$ the chance of acquiring new arcs is independent of age.
The chance of reaching a vertex with a triangle-formation step is
proportional to the degree of the vertex leading to a
\textit{preferential attachment} (an attachment probability
increasing with degree) for high $\beta$ and low $\alpha$. (Note
that the first network model with preferential attachment was a
model of citation networks~\cite{price}.) Fig.~\ref{fig:our} (c)
shows that even though $\beta$ is nearly maximal, the number of
triangles is not as large in the empirical data. The reason for
this is that there are more successful triangle-formation steps
--- or, equally, that it is less probable to attach to a vertex
with lower total degree than the desired total degree of the new
vertex --- for negative $\alpha$. In Fig.~\ref{fig:our}(d) and (e)
we present the results for $\alpha=0$ and $\beta=0$. In this case,
both the aging effect and clustering effect are absent. Not
surprising, neither $T_i$ nor $\lambda_i$ match the real data.
Even though the arcs reach longer back in time for this case, the
number of arcs passing $i$ (i.e.\ $\lambda_i$) is lower. The data
for $\alpha=-1$ and $\beta=0$ are plotted in Fig.~\ref{fig:our}(g)
and (h). We note that with the absence of the triangle-formation
step, not only the number of triangles, but also $\lambda_i$ is
underestimated. As a final comment to Fig.~\ref{fig:our}, the
cusps around $i=21{,}000$ is due to a change in the raw data where
the sampled database was split into different categories and the
sampled papers after this point cites, on average, fewer other
papers.

\begin{figure}
\includegraphics[width=\linewidth]{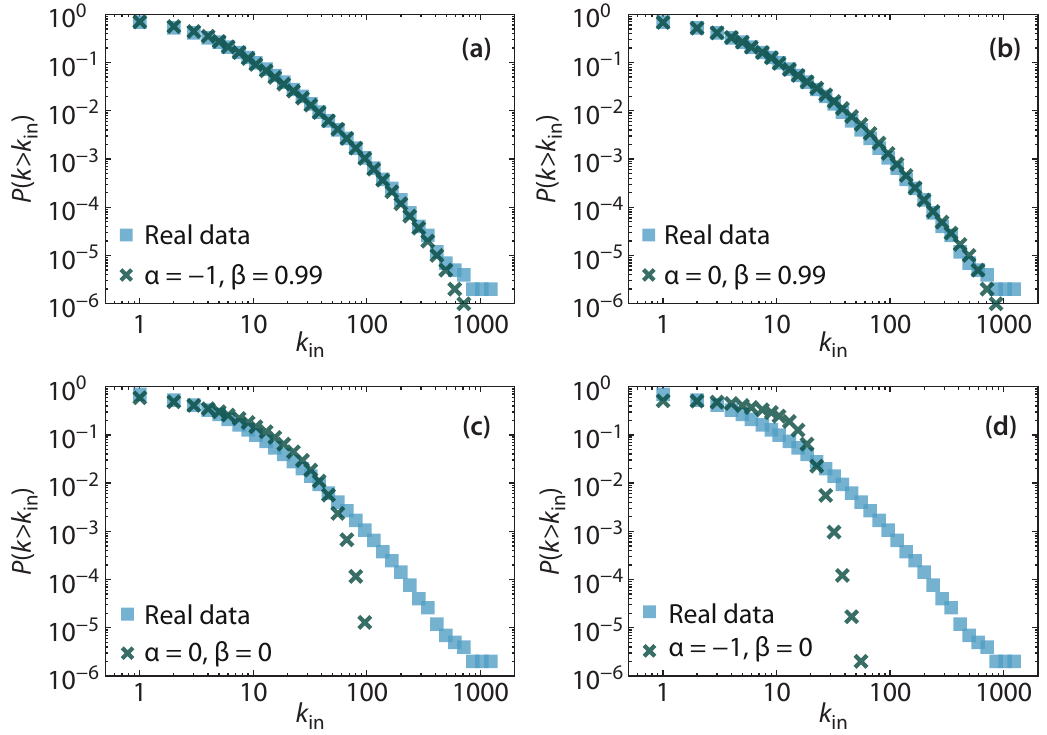}
\caption{(Color online) The in-degree distribution
$P(k_{\mathrm{in}})$ for our four parameter combinations
(indicated in the panels).\label{fig:degdist}}
\end{figure}

Our third quantity is the in-degree distribution that we plot in
Fig.~\ref{fig:degdist}. The both curves with $\beta=0.99$ fit the
real distribution well. As mentioned, there is an effective
(though no necessarily linear) preferential attachment in this
case, which explains the broad distributions. $\lambda_i$ puts
strong constraints on the in degree distribution --- if both
$\lambda_i$ and the out-degree distribution would be fixed to the
observed data (not only the out-degree distribution as in our
case), then the in-degree distribution is the same as the observed
data. With low $\beta$-values, the in-degree distribution becomes
much more narrow than the empirical data. Combining
Figs.~\ref{fig:our} and \ref{fig:degdist}, we note that though
appropriate large value of $\beta$ could generate networks with
in-degree distribution fitting the empirical data, the lacking of
ageing effect would fail to modeling the evolution of citation
network of scientific papers. Taking all these observations into
account, both aging and triangle formation seem to be important
mechanisms in the citation network.

\section{Conclusions}
\label{conclusion} We have proposed a random, evolving network
model for scientific paper citations. In our model, the
attractiveness of a vertex (paper) decays with its age with power
$\alpha$, another parameter $\beta$ determines the number of
triangle formations (when a new paper cites two papers where one
cite the other).  We compared our proposed model with an empirical
citation network of high-energy physics preprints posted at
\url{arxiv.org}. The out-degree distribution is an input to our
model. In this paper we take it from empirical data. We use three
quantities to validate our model --- the number of triangles, the
number of arcs passing the vertex and the degree distribution. All
these quantities are best modeled for parameter values $\alpha=-1$
and $\beta=0.99$~\cite{Notes}. From these observations, our model
suggests that in citation network of scientific papers, the
probabilities of attracting new citations of the papers are about
inversely proportional to their age (measured in its position in
the sequence of publication) and that there is a strong tendency
of citing papers where one paper cites the other. For the future,
we believe it would be informative, as a complement to generative
models like the present, to study the mechanisms of citations by
interview studies and questionnaires to researchers.

\acknowledgments{
This research is supported by the Swedish Research Council
(Z.X.W.) and the Swedish Foundation for Strategic Research (P.H.).}

\bibliographystyle{h-physrev3}

\begin{thebibliography}{99}
\bibitem{Notes}
To fit our model to the data, we scan the region of $\alpha\in[-2,
0]$ and $\beta\in[0, 1]$, with the increment $0.1$ for $\alpha$
and $0.01$ for $\beta$. For each combination of $\alpha$ and
$\beta$, we obtain average $T_i$ and $\lambda_i$ over $200$. After
that, we calculate the goodness-of-fit function $F=\sum_{i=1}^N
|T_i-T^\prime_i|/T^\prime_i+\sum_{i=1}^N
|\lambda_i-\lambda^\prime_i|/\lambda^\prime_i$, where $T_i$ and
$\lambda_i$ are obtained for our model and $T^\prime_i$ and
$\lambda^\prime_i$ for the empirical data. The sum is over each
$200$'th $i$-value (like the analysis of Fig.~\ref{fig:our}). The
combination of $\alpha$ and $\beta$ resulting in the smallest $F$
are our best fit to the empirical data.
\end{thebibliography}

\end{document}